\documentclass[12pt]{article}
\usepackage{pic03}
\usepackage{hyperref}
\usepackage{url}
\usepackage{graphicx}

\begin{document}

\title{
    \bf \textit{KATRIN} - DIRECT MEASUREMENT OF NEUTRINO MASSES IN THE
    SUB-EV REGION
    }
\author{
    Lutz Bornschein for the KATRIN - Collaboration
                                                             \\
        { \em Universit\"at Karlsruhe, }\\
        { \em Institut f\"ur experimentelle Kernphysik am Forschungszentrum Karlsruhe, }\\
        { \em Postfach 3640, 76021 Karlsruhe }
        }

\maketitle

\baselineskip=14.5pt
\begin{abstract}
The \underline{KA}rlsruhe \underline{TRI}tium \underline{N}eutrino
Mass Experiment is a next generation tritium beta decay experiment
designed to reach a sensitivity of $0.2~
\mathrm{eV}/\mathrm{c}^2$. KATRIN will allow to investigate the
role of the neutrino hot dark matter in the evolution of large
scale structures of the universe and will also allow to
discriminate between so-called hierarchical and quasi-degenerated
neutrino mass models. The status of the first components of the
final KATRIN setup will be shown.

\end{abstract}

\baselineskip=17pt

\section{Introduction}

Recent $\nu$-oscillation experiments (e.g. \cite{kamland}) give
compelling evidence that neutrinos have nonzero mass from the
observation of neutrino flavor changes. However, these experiments
are only sensitive to differences between squared neutrino masses.
Neutrinos with masses in the sub-eV range could play an important
role as hot dark matter in the evolution of large scale structures
of the universe. A measurement of the absolute masse scale of
neutrinos could also be decisive in selecting different neutrino
mass models. The most stringent model independent upper limits on
the neutrino mass are given by recent tritium $\beta$ decay
experiments \cite{mainz,troitsk} which are so called "direct mass
measurements". Since these experiments have reached their
sensitivity limit, it is necessary to investigate which
sensitivity could be achieved by a new tritium $\beta$ decay
endpoint experiment (KATRIN).

\section{The KATRIN experiment}

\begin{figure}[t]
  \centerline{\hbox{ \hspace{0.2cm}
    \includegraphics[width=15cm]{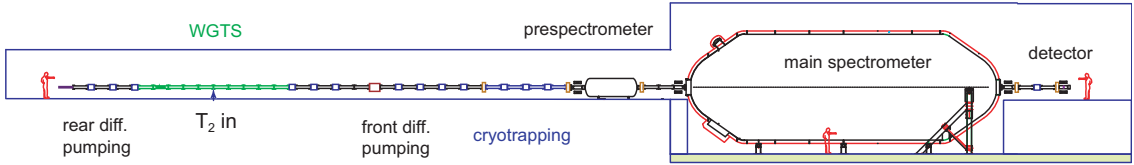}
    }
  }
 \caption{\it
      Sketch of the KATRIN experiment setup (side view).
    \label{katrin} }
\end{figure}

The KATRIN experiment is based on the well proven concept of
recent tritium $\beta$ decay experiments. A sketch of the side
view is shown in Figure \ref{katrin}. The key components are a
\underline{W}indowless \underline{G}aseous molecular
\underline{T}ritium \underline{S}ource (WGTS), differential
pumping sections at the front and rear side, a cryotrapping
section, a system of a pre- and a main spectrometer and a
segmented detector. The KATRIN collaboration has improved the
proposed setup compared to its Letter of Intent \cite{loi}: A
larger source diameter ($90~\mathrm{mm}$) combined with a larger
main spectrometer diameter ($10~\mathrm{m}$), a higher tritium
purity ($95~\%$), an optimized measurement point distribution and
improved systematics. Recent simulations for 3 years of data
taking result in a neutrino mass sensitivity of
$0.2~\mathrm{eV/c^2}~(90\%~\mathrm{C.L.})$ with statistical and
systematic uncertainties contributing about equally. A non-zero
neutrino mass of $0.35~\mathrm{eV/c^2}$ would be detected with a 5
$\sigma$ significance.

\section{The status of first components for the final KATRIN setup}

Exemplary for the ongoing work of the KATRIN setup, the status of
the source and transport system and the pre-spectrometer - two of
the first components for the final KATRIN setup - shall be shown.

\begin{figure}[th]
  \centerline{\hbox{ \hspace{0.2cm}
    \includegraphics[angle=-90,width=15cm]{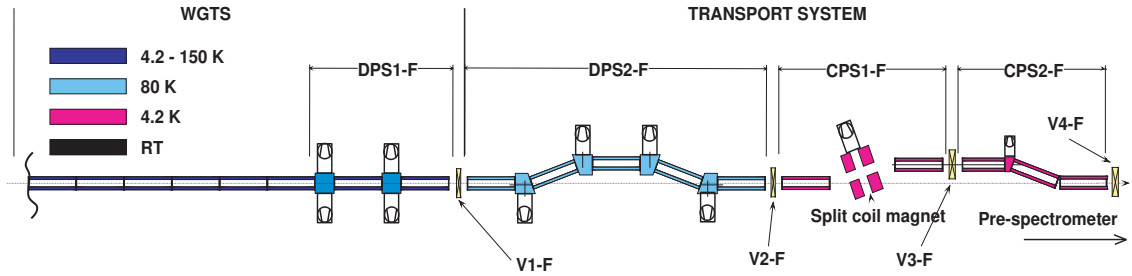}
    }
  }
 \caption{\it
      Reference setup for the KATRIN source and transport system
    \label{source} }
\end{figure}

The WGTS will allow to determine the neutrino mass with a minimum
of systematic uncertainties from the tritium source. The transport
system will guide the $\beta$ decay electrons adiabatically from
the source to the spectrometer, while at the same time eliminating
any tritium flow towards the spectrometer, which has to be kept
practically free of tritium for background reasons. This will be
done by a combination of differential (DPS-F) and cryogenic
(CPS-F) pumping sections (Figure \ref{source}).
 A decision of the reference setup of the source and
transport system for KATRIN has been made in February 2003. The
specification of the DPS2-F has been completed in June 2003 and
the tender action has been started. The specification of the WGTS
and the rest of the transport system is in progress.

The main duty of the pre-spectrometer in the final KATRIN setup is
to reject low energy $\beta$ decay electrons, thus limiting the
number of $\beta$ electrons in the main spectrometer, which
reduces background levels. In the current phase, the
pre-spectrometer will help to verify features of the main
spectrometer design: The vacuum characteristics, the novel concept
of putting the whole spectrometer on retarding potential
(18.6~keV) and the electro magnetic design. For background reasons
the final pressure in the spectrometers has to be below $1 \cdot
10^{-11}~\mathrm{mbar}$ and the outgassing rate has to be below $1
\cdot 10^{-13}~\mathrm{\frac{mbar \cdot l}{s \cdot cm^{2}}}$.
These requirements have recently been met with a vacuum test
chamber. The pre-spectrometer is currently manufactured by the
company SDMS and will be delivered in September 2003. A
substantial test programme will start directly thereafter.

\section{Summary}

KATRIN is a next generation direct neutrino mass experiment with a
sensitivity on the neutrino mass of $0.2~\mathrm{eV/c^2}$. This
allows to distinguish between different neutrino mass models and
to check the role of the neutrinos in structure formation. The
installation of the first components of the final setup is under
way: The reference setup for the source and transport system has
been approved, the tender action for the first parts has been
started. The pre-spectrometer will be delivered in September 2003
and the vacuum requirements have been met with a test chamber.
Tests with the pre-spectrometer concerning the design of the main
spectrometer will start soon.

\end{document}